\documentclass[twocolumn,amsmath,amssymb,aps,pra,floatfix]{revtex4}
\usepackage{bm}
\usepackage{amsfonts}
\usepackage{amsmath}
\usepackage{graphicx}
\usepackage{epstopdf}
\begin{document} 
\title{Guiding neutral polar molecules by electromagnetic vortex field --- a classical approach}
\author{Tomasz Rado\.zycki}
\email{t.radozycki@uksw.edu.pl}
\affiliation{Faculty of Mathematics and Natural Sciences, College of Sciences, Cardinal Stefan Wyszy\'nski University, W\'oycickiego 1/3, 01-938 Warsaw, Poland} 
\begin{abstract}
It is shown, within classical mechanics, that the field of an electromagnetic vortex is capable of capturing and guiding neutral molecules endowed with a permanent electric dipole moment (PEDM). Similarly as in the case of the magnetic field applied to elementary particles or atoms, this effect turns out to be very delicate because of the small values of PEDM observed in real  molecules. They amount to $2\times10^{5}\, e{\mathrm{fm}}$  (electron charge $\times$ fermi) or less, which requires the use of very strong electric fields. It has also been observed that there exists a threshold in field strength above which the particles are ejected from the trap. Trajectories of guided particles are usually quite chaotic, which is a consequence of non-linearity of the equations of motion. With a very special and precise adjustment of parameters, a regular (i.e., circular, in the transverse plane) trajectory can be obtained. The presence of an additional constant electric field pointing along the direction of the wave propagation might help to achieve the necessary tuning and realize such trajectories.

\end{abstract}
%\pacs{} 
\maketitle

\section{Introduction}

Trapping or guiding charged particles due to their interactions with electric or magnetic fields in various configurations constitutes a fairly well-known and well-established phenomenon. Manipulating neutral particles like atoms, molecules or elementary particles like neutron for instance, is much more subtle matter although not new either. This question dates almost $50$ years back to Ashkin's papers \cite{ashkin1,ashkin2} in which the author proposed and reported on experimental accelerating and trapping small neutral particles using the radiation pressure exerted by the laser light. This light pressure was then used to demonstrate the levitation phenomenon of relatively large objects (up to $25 \mu\mathrm{m}$) \cite{ad}, as well as to trap atoms \cite{ashkin3}. 

The other mechanism is based on polarizability of neutral particles, which leads to the appearance of the gradient force as a result of the spatial dependence of the electric field. This kind of traps, called `optical dipole traps' was first considered in \cite{let} and then used to focus the beam of sodium atoms \cite{bjork}. The first atomic trap  --- again for sodium --- based on this technique was reported over 30 years ago in \cite{chu}. Sodium turned out to be here an especially attractive object due to its high electric dipole polarizability at a relatively low mass~\cite{sch}.

Yet another possibility is to use an external non-homogeneous magnetic field to capture neutral particles endowed with non-zero magnetic moment as, for instance, alkali atoms or even neutrons, by creating a certain potential well. In the case of latter ones such a trap was successfully implemented in \cite{kug}, and later for sodium atoms in \cite{mig}.

The idea of a trap being rotated (see for instance~\cite{paul,bbs1}) led in turn to the suggestion of exploiting a rotating electromagnetic wave, i.e., the one that carries a non-zero orbital angular momentum~\cite{ahd,freegarde,slm} and, due to the phase singularity located at the axis of symmetry, called the vortex wave. This idea turned out to be fruitful both for charged~\cite{ibb1} and neutral particles~\cite{ibbtr2}. As in the case of~\cite{kug}, for the latter ones the trapping mechanism was based on the interaction of magnetic moments of particles with the magnetic field of the vortex. Due to very tiny values of magnetic moments of elementary particles, extremely strong fields are needed to trap or guide them but in principle it should be possible~\cite{tr18}. 

In the present paper we would like to apply a similar idea to polar particles, i.e., those endowed with permanent electric dipole moment (PEDM). The symmetries of the Standard Model prevent elementary particles to have PEDM relevant from an experimental point of view~\cite{sm}. Eventual measured non-zero values could be the sign of a `new physics'. On the other hand, there are molecules possessing relatively `large' dipole moments (above $10$ Debyes); among them one can mention the diatomic molecules with ionic bonds like $KBr$, $KCl$, $RbCl$, $BaS$ etc.~\cite{tables}, for which, at least in principle, the effect might be measurable. In a static electric field a molecule rotating with respect to an axis perpendicular to the molecular axis would average its PEDM to zero, unless the external field is strong enough to play the role of a driving force. When a rotating electromagnetic wave is involved, the mutual position between ${\bm E}$ and ${\bm d}$ becomes important, and especially stable trajectories are achieved by synchronizing the motion of these two quantities. This synchronization can be enforced if the wave in question has a sufficiently high intensity.

It should be noted, however, that in typical experimental situations with elementary particles in time dependent electric field, there appears an additional interaction of the magnetic moment with the accompanying magnetic field (such as that given in~(\ref{vEBB})) which turns out to have even stronger influence on the behavior of a particle than the electric one. Then the motion of a particle becomes more complicated as a result of this double interaction. Let us, therefore, compare the impact of these two forces on the motion of molecules dealt with in the present work. Assuming typical magnitude of the magnetic moment of a molecule to be of order of $\mu_B$, the electric dipole moment of order of $10\, {\mathrm D}$ and taking into account that $B/E=1/c$, one gets ${\bm \mu}{\bm B}\sim 10^{-4}{\bm d}{\bm E}$, which means that the magnetic effects may be omitted within our model.

It is quite interesting to observe that the property of trapping particles by a wave carrying orbital angular momentum seems to be quite universal. It is applicable --- as it has already been mentioned --- to both charged and neutral particles provided they exhibit electromagnetic properties. But, interestingly, the same property can be attributed to gravitational waves and, as demonstrated by the early results, leads to trapping massive bodies through a similar mechanism~\cite{ibbgrav}.

Currently electromagnetic waves that exhibit a non-zero orbital angular momentum are generated in many ways such as spiral phase plates~\cite{bei,sueda}, optical fibers~\cite{cz}, computer-generated holograms~\cite{carp} or spatial light modulators~\cite{je}, to mention only a few (for a nice popular introduction see~\cite{padg}). The most commonly used light beams of this category are Bessel-Gauss and Laguerre-Gauss beams, where the names refer to their radial profiles~(see for instance~\cite{ibbzbb,dh}). In the paraxial approximation the electromagnetic fields of both above cases become linear in the radial variable and may be chosen in the simplified form:
\begin{subequations}\label{vEB}
\begin{align}
&{\bm E}({\bm r},t)=\frac{E_0\omega_0}{c}\left[f({\bm r},t),g({\bm r},t),0\right],\label{vEBE}\\
&{\bm B}({\bm r},t)=\frac{E_0\omega_0}{c^2}\left[-g({\bm r},t),f({\bm r},t),0\right],\label{vEBB}
\end{align}
\end{subequations}
with
\begin{subequations}\label{fg}
\begin{align}
&f({\bm r},t)=x\cos\omega_0(t-z/c)+ y\sin\omega_0(t-z/c),\\
&g({\bm r},t)=x\sin\omega_0(t-z/c)-y\cos\omega_0(t-z/c).
\end{align}
\end{subequations}
where $\bm\omega_0$ stands for the wave frequency. As told above, the magnetic field will not play any role here. Consequently, in the non-relativistic approximation, the quotients $z/c$ may be omitted. Then, the electric field satisfies the equation ${\bm \nabla}\times {\bm E}=0$. The motion in the $z$-direction becomes uniform and entirely decouples from the transverse degrees of freedom.

The present paper is organized as follows. In Sect.~\ref{seqmo} classical equations on motions for the translational and rotational degrees of freedom are formulated. From these equations a set of constants of motion together with additional identities can be derived. This is done in Sect.~\ref{com}. Numerical solutions of the equations for certain chosen values of parameters are found in Sect.~\ref{traj}. The appropriate trajectories in the plane perpendicular to the direction of the wave propagation are obtained. It is also analyzed how the values of the parameters of the model (intensity and frequency of the wave, initial energy of particles) affect the motion of particles being guided. In Sect.~\ref{ss} a special circular trajectory (a helix in $3D$) is studied. It is shown that this kind of a stable orbit can be achieved with a precise adjustment of the parameter values. An additional constant electric field oriented along the direction of the wave propagation allows to relax the tuning condition.

\section{Equations of motion}
\label{seqmo}

We will concentrate on molecules or other particles with axial symmetry (with respect to the axis corresponding to the vector of the electric dipole moment $\bm d$), which, in classical mechanics, bears the name of a {\em symmetric top}. For such particles the tensor of inertia $\hat{\bm I}$ may be given the following form:
\begin{eqnarray}
I_{ij}&\!\!\!=&\!\!\!I_\perp\left(\delta_{ij}-\frac{d_id_j}{{d}^2}\right)+I_\parallel\frac{d_id_j}{d^2}\nonumber\\
&\!\!\!=&\!\!\!I_\perp\delta_{ij}-\Delta I\frac{d_id_j}{{d}^2}=I_\perp\left(\delta_{ij}-\kappa \frac{d_id_j}{d^2}\right),
\label{iij}
\end{eqnarray}
where $\kappa=(I_\perp-I_\parallel)/I_\perp=\Delta I/I_\perp$, $d=\sqrt{{\bm d}^2}$. The parameter $\kappa$ is introduced in order to control --- if needed --- the `oblateness' of a given molecule. The values of $\kappa$ close to $1$ correspond to a {\em prolate} (or even {\em linear} in the extreme and most interesting case) particle, $\kappa=0$ refers to a spherical top and negative values describe {\em oblate} ones.

The dynamics of such a molecule results from the interaction between the electric dipole moment and the electric field of the electromagnetic wave. The following equations of motion constitute the complete set for translational and rotational degrees of freedom of a molecule of mass $m$:
\begin{subequations}\label{eqmot}
\begin{align}
&m\frac{d^2}{dt^2}{\bm r}=({\bm d}\cdot {\bm\nabla}){\bm E},\label{eqmotr}\\
&\frac{d}{dt}{\bm J}={\bm d}\times {\bm E},\label{eqmotj}\\
&\frac{d}{dt}{\bm d}= {\bm \omega}\times {\bm d}.\label{eqmotd}
\end{align}
\end{subequations}
where $\bm r$ denotes its position and $\bm J$ the angular momentum. The latter can be written as 
\begin{eqnarray}
{\bm J}=\hat{\bm I}{\bm \omega}=I_\perp\left({\bm \omega}-\kappa\,\frac{({\bm \omega}\cdot{\bm d}){\bm d}}{d^2}\right)=I_\perp{\bm \omega_\kappa},
\label{io}
\end{eqnarray}
where we have introduced the `projected' angular velocity ${\bm \omega_\kappa}={\bm \omega}-\kappa({\bm \omega}\cdot{\bm d}){\bm d}/d^2$ in place of the true angular velocity $\bm \omega$ connected with the rotation of the molecule. This leads to some simplification of the equations of motion:  
\begin{subequations}\label{eqmo}
\begin{align}
&m\frac{d^2}{dt^2}{\bm r}=({\bm d}\cdot {\bm\nabla}){\bm E},\label{eqmor}\\
&I_\perp\frac{d}{dt}{\bm \omega_\kappa}={\bm d}\times {\bm E},\label{eqmoj}\\
&\frac{d}{dt}{\bm d}= {\bm \omega_\kappa}\times {\bm d}.\label{eqmod}
\end{align}
\end{subequations}
In fact, when guiding a polar particle along a beam of radiation, we are not interested in the value of $\bm \omega$, but mainly in the position $\bm r$. Therefore, $\bm \omega_\kappa$ is a quantity equally good in our considerations as $\bm \omega$. Apparently $\bm \omega_\kappa$ might be (at least in principle) eliminated from the above equations, resulting in the motion independent of the value of $\kappa$. This is not true, since $\kappa$ enters through the initial conditions for $\bm\omega$ and  $\bm d$. Nonetheless from~(\ref{eqmo}) one can draw a conclusion that for any `oblateness' the same trajectory can be obtained by means of the appropriate modification on the initial conditions, although the rotational states will be different. Equations containing $\bm \omega$ are much more intricate because of the nontrivial time dependence of the moment of inertia components connected with the instantaneous orientation of the electric dipole moment.

The set of equations~(\ref{eqmo}) is highly nonlinear and it is unlikely to have it solved in an exact, analytical way, maybe apart from some special cases. General trajectories of a guided molecule are mainly obtained through the numerical solutions of the equations. For the numerical analysis it is convenient to introduce dimensionless quantities similarly as it was done in~\cite{ibbtr2}:
\begin{subequations}\label{para}
\begin{align}
&{\bm \xi}=k{\bm r},\;\;\;\; {\bm \eta}=\frac{\bm d}{d}, \;\;\;\;\beta=\frac{E_0 d}{mc^2},\;\;\;\;\alpha=\frac{E_0 d}{I_\perp \omega_0^2}, \label{para1}\\
&\tau=\omega_0 t,\;\;\;\; {\bm \Omega}_\kappa=\frac{{\bm \omega}_\kappa}{\omega_0},\;\;\;\; {\bm \Omega}=\frac{\bm \omega}{\omega_0}.\label{para2}
\end{align}
\end{subequations}
where $k=\omega_0/c$.

Using these quantities and appropriately rewritten electric field~(\ref{vEBE}), together with~(\ref{fg}) taken in the non-relativistic approximation, the equations of motions can be given the form (here $\cdot=d/d\tau$):
\begin{subequations}\label{dx}
\begin{align}
&\ddot{\bm \xi}=\beta {\bm {\mathcal E}}({\bm \eta},\tau),\label{dxi}\\
&\dot{\bm \Omega}_\kappa=\alpha {\bm \eta}\times {\bm {\mathcal E}}({\bm \xi},\tau)\label{dom}\\
&\hspace{5ex}=\alpha[\eta_z(\xi_y\cos\tau-\xi_x\sin\tau), \eta_z(\xi_x\cos\tau+\xi_y\sin\tau),\nonumber\\
&\hspace{8ex}(\eta_x\xi_x-\eta_y\xi_y)\sin\tau-(\eta_x\xi_y+\eta_y\xi_x)\cos\tau],\nonumber\\
&\dot{\bm \eta}={\bm \Omega}_\kappa\times {\bm \eta},\label{deta}
\end{align}
\end{subequations}
where
\begin{equation}
{\bm {\mathcal E}}({\bm \rho},\tau)=[\rho_x\cos\tau+\rho_y\sin\tau,\rho_x\sin\tau-\rho_y\cos\tau,0].
\label{Ecal}
\end{equation}

\begin{figure}[h]
\begin{center}
\includegraphics[width=0.45\textwidth,angle=0]{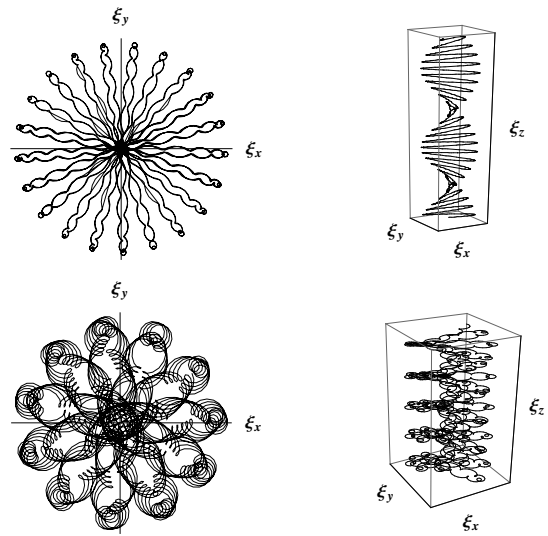}
\end{center}
\caption{The exemplary two- and three-dimensional trajectories of a neutral, polar particle in an electromagnetic vortex field. On the left the motion in the perpendicular plane and on the right the corresponding trajectory in $3D$ are depicted.}
\label{example}
\end{figure}

As mentioned, this is the complicated, nonlinear set for nine unknown functions, the general solution of which is not available. In Figure~\ref{example} certain exemplary numerical solutions, for arbitrarily chosen values of the constants $\alpha$ and $\beta$ are presented in the graphical form. 

The motion in the direction of the wave propagation (the $z$-direction) is free. Contrary to this, in the perpendicular plane ($xy$) the motion is bounded, thus providing the opportunity to guide particles or molecules along a beam of light. In the following sections the set of equations~(\ref{dx}) will be studied in more detail.

\section{Constants of motion and identities}
\label{com}

Manipulating the equations of motion, one can derive a couple of formulas for the constants of motion. The first two are rather obvious. They are the velocity of the motion along the $z$-axis and the absolute value of the electric dipole moment:
\begin{subequations}\label{constHa}
\begin{align}
&{\cal H}_1=\dot{\xi}_z,\label{constH1}\\
&{\cal H}_2={\bm \eta}^2.\label{constH2}
\end{align}
\end{subequations}
It should be noted, that $\dot{\xi}_z$ might not be a constant if relativistic effects were taken into account. This phenomenon was observed and derived in~\cite{ibb1} for charged particles moving in an electromagnetic vortex field and later in combination with a constant magnetic field~\cite{ibbtrloc}. However, the model considered in the present paper is non-relativistic by definition since the electric dipole moment is not a relativistic concept.

The following constant of motion, i.e.,
\begin{equation}
{\cal H}_3=\frac{1}{2\beta}\dot{\bm \xi}^2+\frac{1}{2\alpha}{\bm \Omega}_\kappa^2-{\bm \eta}\cdot {\bm {\mathcal E}}({\bm \xi},\tau)-\frac{1}{\alpha}\Omega_{\kappa\, z},\label{constH3}
\end{equation}
constitutes {\em de facto} an implementation of the principle of energy conservation. Strictly speaking, the energy in the considered system is not conserved due to the time dependence of the electric field of the vortex, which is not an ingredient of a dynamical system but rather an external field remaining under our control. This is reflected by the presence of the last term in~(\ref{constH3}). The first three terms are just the kinetic energy related to the translational and rotational degrees of freedom and the potential energy of the dipole interaction with the electric field. On the other hand from~(\ref{dom}) one obtains
\begin{eqnarray}
\frac{d}{d\tau}\left(\frac{1}{\alpha}\Omega_{\kappa\, z}\right)&\!\!\!=&\!\!\!(\eta_x\xi_x-\eta_y\xi_y)\sin\tau-(\eta_x\xi_y+\eta_x\xi_y)\cos\tau\nonumber\\
&\!\!\!=&\!\!\!\frac{\partial}{\partial\tau}\left(-{\bm \eta}\cdot {\bm {\mathcal E}}({\bm \xi},\tau)\right),
\label{po}
\end{eqnarray}
i.e., the variation in potential energy due to the explicit time evolution of the electric field. This term constitutes, therefore, the external source of the energy for the system.

The subsequent constants of motion are the $z$-component of the angular momentum stemming both from the translational and rotational motions:
\begin{equation}
{\cal H}_4=\frac{1}{\beta}(\xi_x\dot{\xi}_y-\xi_y\dot{\xi}_x)-\frac{1}{\alpha}\Omega_{\kappa\, z}\label{constH4}
\end{equation}
and the projection of the angular velocity of the rotating particle onto the direction of the electric dipole moment, which is a direct consequence of the equations~(\ref{dom}) and~(\ref{deta}):
\begin{equation}
{\cal H}_5={\bm \eta}\cdot {\bm \Omega}_\kappa=(1-\kappa){\bm \eta}\cdot {\bm \Omega}.\label{constH5}
\end{equation}
This reflects the obvious fact, that the electric field is unable to modify the component of the angular momentum of the rotating particle parallel do $\bm d$. The constant ${\cal H}_5$ becomes trivial for linear molecules since then this component identically vanishes.

One can also derive several identities useful for analyzing the motion: 
\begin{subequations}\label{iden}
\begin{align}
&\dot{\bm \eta}^2-{\bm \Omega}_\kappa^2={\cal H}_5^2,\label{iden1}\\
&\ddot{\bm \xi}^2=\beta^2(\eta_x^2+\eta_y^2),\label{iden2}\\
&\dot{\Omega}_{\kappa\, x}^2+\dot{\Omega}_{\kappa\, y}^2=\alpha^2\eta_z^2(\xi_x^2+\xi_y^2),\label{iden3}
\end{align}
\end{subequations}
where the first one can be proved to be equal to the square of~(\ref{constH5}).

In particular, for a uniform circular trajectory, for which the (centripetal) acceleration is constant in time, one can deduce from (\ref{iden2}) that the dipole has to perform a uniform rotation with respect to the $z$-axis. This is confirmed by a direct calculation in section~\ref{ss}.

\begin{figure*}[t]
\begin{center}
\includegraphics[width=0.95\textwidth,angle=0]{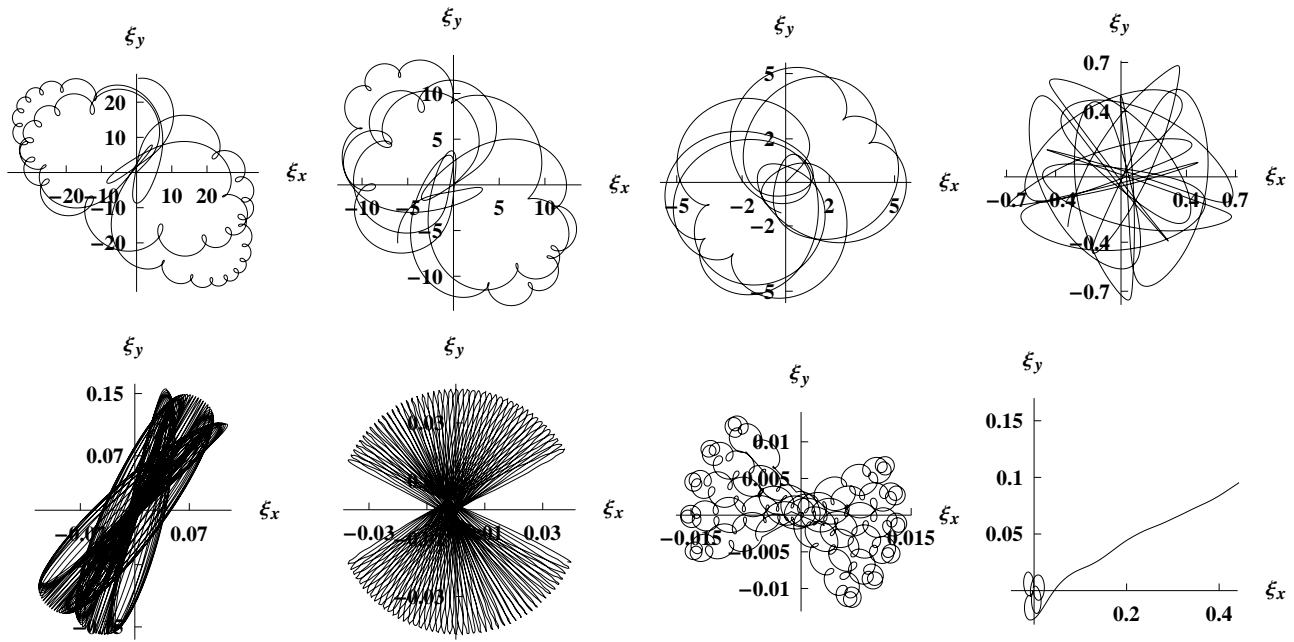}
\end{center}
\caption{The modification of the particle trajectory with increasing field strength. The subsequent values of parameters are: $\gamma=10^{-10},10^{-9},10^{-8},10^{-7},10^{-6},10^{-5},4 \times 10^{-4},4\times 10^{-3}$. It is visible that for more intense fields the trajectories become tighter. If some threshold value of the electric field is exceeded, the trajectory becomes unstable. The unit on the axes is the wavelength $\lambda$.}
\label{gamma}
\end{figure*}

\section{Trajectories}
\label{traj}

In this section we would like to draw and analyze numerical solutions of the equations of motion in the perpendicular plane. In general, the effects exerted by the vortex field on neutral particles are very delicate and in practical applications it is necessary either to pre-cool particles to be trapped or to use extremely powerful fields. Therefore, in the following subsection, we will focus on the strength of the electromagnetic field and initial kinetic energy of the particles and try to establish the magnitudes of these quantities for which the trapping ability of the vortex field can be achieved. The size of traps is of interest as well.

As it has already been mentioned, the `oblaticity' parameter $\kappa$ does not enter into the equations of motion~(\ref{dx}). It is hidden behind the quantity $\Omega_\kappa$ and solutions of the equations for $\xi$ and $\eta$ do not depend on its value. It intervenes only through the initial state of particle rotation, and therefore, it does not have to be independently studied.  

\subsection{Field strength}
\label{gam}

A glance at the equations~(\ref{dx}) shows that after rescaling the vector $\bm\xi$ with the factor $\beta$, the only parameter that remains is $\gamma=\alpha\,\beta$ which is proportional to the wave intensity or electric field squared. It should be remembered, of course, that the initial conditions for the particle position or energy are rescaled in a similar manner (and hence, the parameter $\beta$ will still be present in the solutions), which should be taken into account when solving the problem numerically and scaling the axes of the coordinate system. 

The actual values of PEDMs of neutral molecules, atoms or particles are extraordinarily small. Relatively `large' value is measured for the $\Lambda^0$ hyperon for which $d_\Lambda=(-3.0\pm 7.4)\times 10^{-4}\, e\, \mathrm{fm}$~\cite{lambda}. For the neutron the observed limit is much lower $d_n\lesssim (-0.21\pm 1.82)\times 10^{-13}\, e\, \mathrm{fm}$~\cite{neutron} at roughly comparable masses of both particles.

In the case atoms, the observed upper limits account for $10^{-11} - 10^{-9}\, e\, \mathrm{fm}$~(for instance~\cite{cs,rad,play}, see also the review article by Chupp {\em et al.}~\cite{chupp}). As to the molecules the measurements show that PEDMs reach the value of $2\times 10^{5}\, e\, \mathrm{fm}$ for some linear ones. As an example can serve here the diatomic molecules of $KBr$, $KCl$, $RbCl$ or $BaS$~\cite{tables}.
Estimated moments of inertia account for from $2.6\times 10^{-3}\, \mathrm{ u\, fm^2}$ for the neutron (and not very different for $\Lambda^0$) to $6\times 10^9\, \mathrm{ u\, fm^2}$ for small molecules.

Due to the extremely small values of PEDM the trapping or guiding the particles due to the mechanism described in this work has by now rather theoretical nature and is limited to molecules since very intense fields would be required in experiments to achieve the goal. For instance electric fields of order of $100\, \mathrm{kV/cm}$ (which is, however,  still much lower than the internal field in a molecule) yield the values of the parameter $\gamma$ falling within the range from $2.5\times 10^{-28}$ for $\Lambda^0$ to $10^{-12}$ for molecules (assuming the electromagnetic wave frequency of order of $\omega_0\approx 10^{12}\, {\mathrm s}^{-1}$). This is the typical situation when trapping and guiding neutral particles, that the effects are very tiny: this equally refers  to the particles endowed with magnetic moment~\cite{ibbtr2,tr18} in interaction with magnetic field. 

Figure~\ref{gamma} presents the numerical solutions of the equations of motion in the form of the particle trajectories for eight increasing values of $\gamma$. The trap would still be effective even for the values of $\gamma$ which are several of orders of magnitude smaller than those presented in the figure but the trajectories become then very extended, and the time of particles oscillations very large. The fact that the trap would still play its role is due to its spatial extent (in fact unlimited in our model) and to the linear growth of the field in radial direction, according to the formulas~(\ref{vEB}) and (\ref{fg}). In real circumstances, the beam of electromagnetic field is not so wide, and in any case, the range of applicability of the paraxial approximation is limited in space to distances of order of $\lambda$~\cite{ibbtrloc}. Roughly, one can say that the trajectories plotted in first three pictures correspond to escaping particles, and only a high increase of the electric field strength entails the capture of particles, as in the following four plots.
 
\begin{figure}[h]
\begin{center}
\includegraphics[width=0.35\textwidth,angle=0]{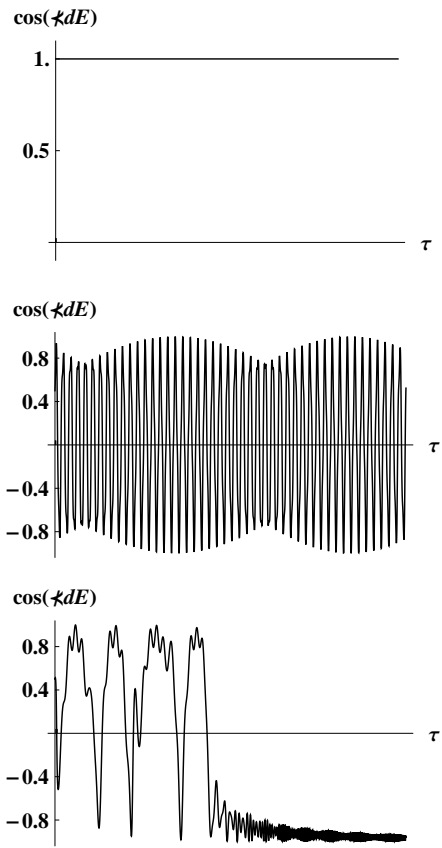}
\end{center}
\caption{The comparison of the cosine of the angle between temporary directions of the PEDM and of electric field creating a vortex. The last plot shows their (permanent) almost anti-parallel orientation when kicking the particle out of the field.}
\label{etaE}
\end{figure}

The interesting effect to be mentioned is the breakdown of the trap when the vortex field becomes too strong. This is seen on the last plot, where a particle, after a couple of oscillations, is kicked off the electromagnetic field. The mechanism of this effect is explained in the third plot of Figure~\ref{etaE}. Contrary to the first two drawings, where PEDM is either rotating constantly parallel to the electric field (as in the case of circular motion dealt with in Sec.~\ref{ss}) or oscillating with vanishing mean value (as in the case of more chaotic motion), in the unstable case the PEDM is captured in an anti-parallel orientation. This results in the appearance of the radial force pointing outwards, i.e. towards the larger electric field, instead of the vortex axis. As a result, the particle is kicked out of the field. This effect occurs in magnetic traps as well~\cite{paul}. It is so, since, for a very strong field, the greatest drop in potential energy in the four-dimensional space (the space of two angles of $\bm \eta$, $\xi_x$ and $\xi_y$) for an anti-parallel orientation is achieved not by the rotation of the dipole, but by its ejection from the field. This phenomenon can be eventually compensated for by increasing the frequency of electromagnetic wave and consequently, the frequency of the rotation of $\bm E$.

One should mention the observed fact of high sensitivity of the shape of trajectories on the initial orientation of the PEDM with respect to the electric field. It is a natural consequence of the highly nonlinear character of the differential equations~(\ref{dx}), which property introduces some chaotic character to the motion.

\subsection{Initial energy}
\label{vel}

\begin{figure*}[t]
\begin{center}
\includegraphics[width=0.9\textwidth,angle=0]{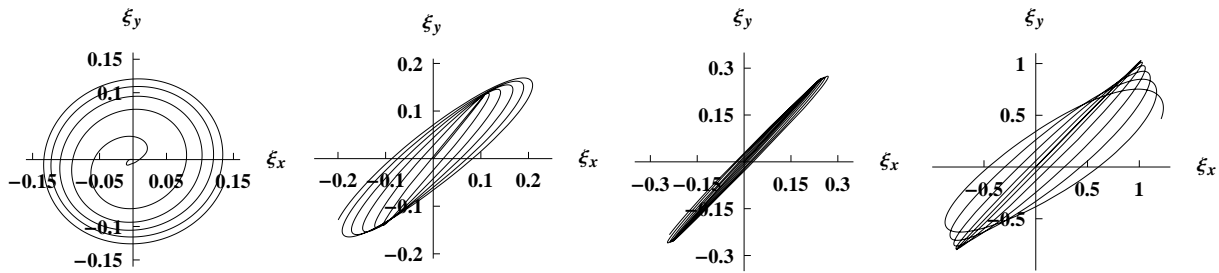}
\end{center}
\caption{Widening of trajectories with increasing initial kinetic energy of captured particles. In these plots $\gamma = 10^{-10}$ and initial velocities (chosen in the radial direction) are subsequently: $10^{-7} c, 10^{-5} c, 2\times 10^{-5}c,7\times 10^{-5}c$.}
\label{velocity}
\end{figure*}

Quite a common situation in trapping neutral particles, whether by electric or magnetic field, is the need to cool them down to keep them trapped~(see for instance \cite{adams,ash,balykin}). Due to the relatively shallow potential wells that are obtained in realistic situations, cooling down to millikelvins or even microkelvins is required. This can be achieved for instance by laser cooling and partially in a simple way of letting hot particles to escape from the trap. The same happens in the guiding trap currently under consideration, implemented by the interaction of the PEDM with the electric field of the vortex. The escape of particles caused by their excessive energies is a result of the finite size of the trap, which in our model (paraxial approximation) is, formally, not feasible but nevertheless, can be visualized. 
Figure~\ref{velocity} shows the gradual expansion of the trajectory (up to the dimensions exceeding the trap size) with the increase of the initial kinetic energy of the particles (all drawings cover the same time interval). This might be expected, but is not obvious due to the strongly chaotic nature of the equations of motion. The fact that the escape phenomenon occurs only for very high velocities is due to extremely strong fields ($\gamma=10^{-10}$), and therefore, the plots should be treated as illustrative rather than quantitative. It can be roughly estimated, however, that for a particle with PEDM of $10\,\mathrm{D}$, with transverse dimensions of the trap of order of the wavelength and an electric field of about $1\, \mathrm{kV/cm}$, particles should be cooled down to temperatures of few millikelvins, which is quite a typical result for the very shallow binding potential.

Of course, apart from rather regular trajectories shown in the figure, there are also many more chaotic trajectories revealing the same phenomenon.

As we already know, in the case of extremely strong fields, formally capable of trapping particles, there appears another escape mechanism, mentioned in the previous sub-section (see last plots in Figures~\ref{gamma} and \ref{etaE}).  
It should also be added that in quantum picture, the escape can also be realized through tunneling effect~\cite{tr18}.

\section{Circular solution}
\label{ss}

Among all trajectories, more or less chaotic in nature, one can obtain, upon fine tuning of parameters a certain regular, circular trajectory. In order to find out that it is really possible let us look for a solution in the following form:
\begin{equation}
\xi_x(\tau)=\xi_0 \cos \sigma\tau,\;\;\;\; \xi_y(\tau)=\xi_0 \sin \sigma\tau,
\label{circxi}
\end{equation}
where $\xi_0$ and $\sigma$ are certain constants. Then, from~(\ref{dxi}) one easily obtains
\begin{equation}
\eta_x(\tau)=\eta_0 \cos (1-\sigma)\tau,\;\;\;\; \eta_y(\tau)=\eta_0 \sin (1-\sigma)\tau,
\label{circeta}
\end{equation}
with $\eta_0=-\sigma^2\xi_0/\beta\leq 1$ and $\eta_{z0}=\pm \sqrt{1-\eta_0^2}$. If such a solution exists, it can be easily verified that
\begin{equation}
{\bm \eta}\cdot {\bm {\mathcal E}}({\bm \xi},\tau)={\bm \xi}_0\cdot {\bm \eta_0}
\label{cos}
\end{equation}
i.e., the projection of the PEDM onto the electric field is time independent since these vectors stay constantly parallel, in agreement with the first plot of Figure~\ref{etaE}.

Integrating~(\ref{dom}), one gets
\begin{equation}
{\bm \Omega}_\kappa=\frac{\alpha\eta_{z0}\xi_0}{1-\sigma}[\cos(1-\sigma)\tau,\sin(1-\sigma)\tau,0]+{\bm \Omega}_{\kappa 0}.
\label{circom}
\end{equation}
Now it stems from~(\ref{deta}) that the integration constant(s) ${\bm \Omega}_{\kappa\, 0}$ for this type of a motion to be realized, must have the form:
\begin{equation}
{\bm \Omega}_{\kappa 0}=\left[0,0, 1-\sigma+\frac{\alpha(1-\eta_0^2)\xi_0}{\eta_0(1-\sigma)}\right],
\label{okz}
\end{equation}
which means that at the initial moment the PEDM has to perform the precession around the direction of the propagating wave. This precession is then preserved throughout the motion.

The fine-tuning condition
\begin{equation}
(1-\sigma)^2-\frac{\gamma\eta_{z0}^2}{\sigma^2}-\Omega_{\kappa z0}(1-\sigma)=0,
\label{ft1}
\end{equation}
can be released by inserting into the system a fixed electric field oriented along the z-axis (${\bm E}_z$). Such a field of adequate value enforces the desired precession leading to a condition easier to satisfy
\begin{equation}
(1-\sigma)^2-\frac{\gamma\eta_{z0}^2}{\sigma^2}-\Omega_{\kappa z0}(1-\sigma)=\eta_{z0} \alpha_z,
\label{ft2}
\end{equation} 
where
\begin{equation}
\alpha_z=\frac{E_z d}{I_\perp \omega_0^2}.
\label{ezc}
\end{equation}

This situation is somewhat similar to that of~\cite{ibbtr2} where a certain constant magnetic field parallel to the $z$ axis was shown to possess a resonant values, for which spinning particles were carrying out a circular motion. In the case of charged particles, in turn, the combined action of the vortex field and a constant magnetic field caused the pinning of Landau's orbits~\cite{ibbtrloc}. Surely, one controls neither $\eta_{z0}$ nor $\Omega_{\kappa z0}$, but particles not satisfying~(\ref{ft1}) or (\ref{ft2}) finally escape from the trap, leaving inside only those performing the stable motion~(\ref{circxi}).

Keeping in mind that $\gamma$ is in general very small, the resonance condition~(\ref{ft2}) can be given the elegant form:
\begin{equation}
I_\perp(\omega_0-\omega_r)^2={\bm E}_z\cdot {\bm d}, 
\label{rescon}
\end{equation}
where $\omega_r=\sigma \omega_0$ is the angular velocity of the particle in a circular motion state. This equation may be called a {\em viral relation} because it states that the kinetic energy in the rotational motion of a solid (this solid on the circular trajectory must be rotating with velocity equal to $\omega_0-\omega_r$ in order to stay synchronized with the vortex field) amounts to minus one half of the potential energy of the interaction of a dipole with the external field.

\begin{figure}[h]
\begin{center}
\includegraphics[width=0.45\textwidth,angle=0]{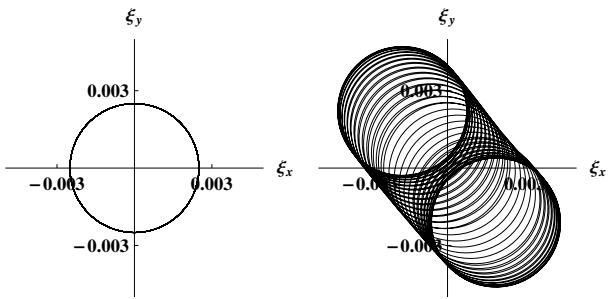}
\end{center}
\caption{Circular trajectory for fine-tuned parameters on the left and for slightly detuned ones on the right.}
\label{circlesym}
\end{figure}

As it is obvious, the solution~(\ref{circxi}) satisfying~(\ref{ft1}) corresponds to a certain circular trajectory shown on the left plot in Figure~\ref{circlesym}. For the parameters slightly detuned the trajectory gets changed into one (obtained numerically from~(\ref{dx})) for which the circle travels back and forth, retaining its character, as depicted on the right plot. 

\begin{figure}[t]
\begin{center}
\includegraphics[width=0.45\textwidth,angle=0]{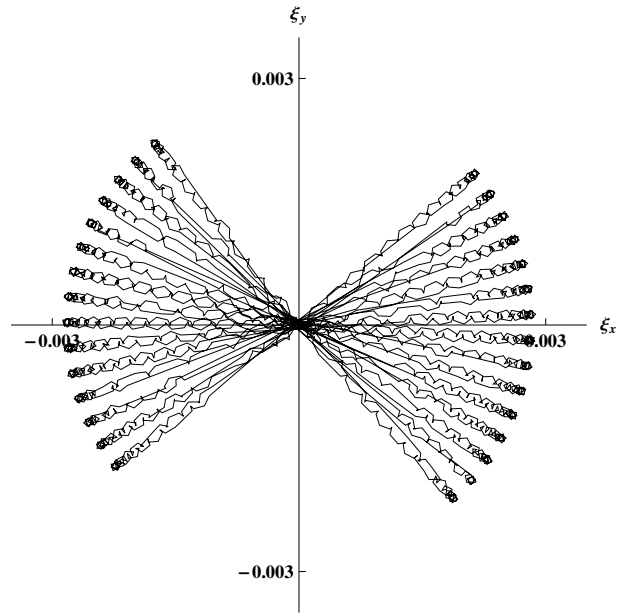}
\end{center}
\caption{Motion of the time-averaged position (i.e., the center of the moving circle), for the slightly detuned parameters as in Fig.~\ref{circlesym}, but for much larger time.}
\label{centermotion}
\end{figure}

What is interesting, for still longer times $\tau$, the circular orbit travels across the plane, still preserving the circular shape.  
In order to better visualize this kind of a motion without losing transparency, Figure~\ref{centermotion} was performed to show the traveling position of the center of this circle when displacing. The line is obtained by averaging the particle position over the time of a single cycle.
 
\section{Summary}
\label{sum}

The current work was concerned with the propagation of neutral particles (mainly molecules), endowed with the permanent electric dipole moment in the field of an electromagnetic wave with phase singularity. It has been shown that this kind of a wave is able to trap such neutral objects and guide them along the vortex core. Due to the very tiny values of the electrical moments, this can be achieved by using huge values of the external fields. This is quite typical of magnetic traps, as in general the potential depth resulting from the interaction of magnetic moment and magnetic field is very shallow. This, naturally, requires the adequate pre-cooling of particles. A similar situation occurs in the current setup. It was shown, in a numerical manner, that the cooling of particles helps to prolong their trapping time.

A mechanism of ejecting the particles from the very strong field has been revealed. It is connected with the anti-parallel positioning of the PEDM and the rotating electric field, where it is energetically more beneficial to eject the particle outside than to turn the dipole moment. 

The obtained trajectories exhibit in general chaotic behavior. This has been observed for particles guided or trapped via the magnetic moment interactions and is a consequence of strongly nonlinear nature of the equations of motion. The parameters can be, however, fine tuned so that the obtained trajectory becomes regular and stable: the particle then runs along a helix. Introducing an additional constant electric field along the vortex core allows to better control the motion and extend the range of parameters for which the motion is circular.

An interesting extension of the present work, with more practical realizations, would be to investigate the same mechanism of trapping and guiding neutral particles without PEDM but with a dipole moment induced by a vortex field.

\end{document}